\documentstyle[11pt,newpasp,twoside,epsf]{article}
\def\edcomment#1{\iffalse\marginpar{\raggedright\sl#1\/}\else\relax\fi}
\reversemarginpar

\begin{document}
\title{A steeper stellar mass spectrum in the Outer Galaxy?}
 \author{J. Brand}
\affil{Istituto di Radioastronomia - C.N.R., Via Gobetti 101, I-40129 Bologna,
Italy (brand@ira.cnr.it)}
\author{J.G.A. Wouterloot}
\affil{Joint Astronomy Center, Hilo, Hawaii, USA}
\author{A.L. Rudolph}
\affil{Dept. of Physics, Harvey Mudd College, Claremont, CA, USA}
\author{E.J. de Geus}
\affil{Neth. Found. for Research in Astronomy, Dwingeloo, The Netherlands}

\begin{abstract}
We discuss the results of high-resolution ($\sim 0.1-0.2$~pc) BIMA CO 
observations of the central regions of 3 molecular clouds in the far-outer 
Galaxy (FOG). We identify clumps and investigate their stability by using the 
virial theorem, including
terms due to gravity, turbulence, magnetic field, and interclump gas pressure,
and make a comparison with clumps in local clouds (RMC and Orion B South).
While a reasonable combination of these forces can render most clumps stable,
an interesting difference between FOG and local clumps emerges when comparing
only gravity and turbulence. In the FOG these forces are in equilibrium
(virial parameter $\alpha \approx 1$) for clumps down to the lowest masses
found (a few M$_{\odot}$), but for local clumps $\alpha \approx 1$ only for
clumps with masses larger than a few tens of M$_{\odot}$. Thus it appears that
in the FOG gravity is the dominant force down to a much lower mass than in
local clouds, implying that gravitational collapse and star formation may occur more readily even
in the smallest clumps. This might explain the apparently steeper IMF found in
the outer Galaxy.
\end{abstract}

\section{Introduction}
These observations are part of a long-term study in which, through a
comparison of molecular clouds across the Galaxy, we aim to find the
influence of a different physical environment on the properties of molecular
clouds and their embedded star formation. Previously, we have studied
molecular clouds at galactocentric distances $R > 15$~kpc (Brand \& Wouterloot
1994) and analyzed molecular cloud properties across the Galaxy (Brand \&
Wouterloot 1995), using single-dish observations. Using the far-infrared
luminosities of IRAS point sources in the outer Galaxy, Wouterloot et al.
(1995) derived a slope for the IMF which is {\it steeper} than that measured
in the solar neighbourhood. In order to investigate this further we need to
study the properties of clumps (sizes from $0.2-2$~pc) in molecular clouds in 
the outer Galaxy, and higher-resolution observations are required. Such
observations are presented here. 

\section{Far-outer Galaxy clouds}
Salient details of the three clouds we have observed in CO(1--0) 
are listed in Table~1. 
They were originally found in a single-pointing CO survey of IRAS
sources in the outer Galaxy (Wouterloot \& Brand 1989; WB89). In each cloud, 
we mapped the region around the IRAS source with BIMA, while the whole cloud
was mapped in CO(1--0) (WB89-85; NRAO 12-m) or CO(2--1) (WB89-380, -437; KOSMA 
3-m), to provide the zero-spacing baselines and mass estimates. 
The available data (radio-continuum, water masers, outflows) indicate that 
WB89-437 is the youngest star-forming region of the three, and WB89-85 the 
oldest.

\begin{table}
\caption{The three far-outer Galaxy clouds observed}
\begin{tabular}{rcrlll}
\tableline
Name & R & \multicolumn{1}{c}{d} & M$_{\rm CO}$ &
L$_{\rm fir}$ & Associated radio emission\\
WB89- & \multicolumn{2}{c}{(kpc)} & (10$^4$~M$_{\odot}$) &
(10$^4$~L$_{\odot}$) & \\
\tableline
85  & 15.0 & 11.5 & 1.6 & 10 & Evolved (Sh~2-127) + compact; \\
    &      &      &     &    & both optically visible\\
380 & 16.6 & 10.3 & 3.5 & 10 & Compact; optically invisible\\
437 & 16.2 & 9.1 & 0.94 & 7.1 & None. Opt. visible H{\sc ii} region\\ 
    &      &      &     &    & in same cloud (WB89-436) \\
\tableline
\tableline
\end{tabular}
\end{table}

\noindent
The data were analyzed using the 3-D clump detection and analysis
program CLUMPFIND (Williams, de Geus, \& Blitz 1994). The program assigns 
virtually all emission in the BIMA maps (above a 2.5$\sigma$ 
threshold) to clumps. The clump lists were trimmed to include only resolved 
clumps, that lie completely inside the map boundary.

\subsection{Local comparison sample}
The physical parameters of the clumps found in the FOG clouds were compared
to those of clumps in two local clouds: the Rosette
Molecular Cloud (RMC: Williams, Blitz, \& Stark 1995, $^{13}$CO(1--0) at 0.8~pc
resolution; Schneider et al. 1998, $^{13}$CO(2--1) at 0.12~pc resolution)
and Orion B South (Kramer, Stutzki, \& Winnewisser 1996, $^{13}$CO(2--1) at 
0.14~pc resolution).
The original data were provided by the authors, and on both clouds we
performed the same clump analysis as on the FOG clouds. 

\section{Clump stability}
A simplified picture of a molecular cloud is that of 
an ensemble of relatively high-density clumps, moving
about in a low-density (interclump) medium. Various forces are at work on
the clumps: their (self-) gravity and the pressure of the interclump medium
try to compress the clumps, while the pressures due to turbulent and thermal
motions, and due to magnetic fields work in the opposing direction. The
virial theorem for a clump, with external pressure $P$, can therefore be
written as 

\noindent
$\rm P/k = P_{turb}/k + P_{grav}/k + P_{magn}/k =
\bar \rho \sigma^2 /k - G M_{CO} \bar \rho /3 r k + B^2 /8  \pi k$,
where the P$_{\rm turb}$ includes both thermal and (dominant) non-thermal
contributions, and with $\sigma$ the 1-D velocity dispersion, $\bar \rho$ the
average mass density, and M$_{\rm CO}$, r the clump mass and radius.
The ratio between turbulent and gravitational pressures is the virial
parameter $\alpha$.
In Fig.~1 we plot $\alpha$ and various pressure ratios as a function of clump
mass. If a clump is in virial equilibrium with external surface pressure
$P$=0 and in the absence of a magnetic field, the turbulent pressure is
exactly balanced by the gravitational pressure, and $\alpha$=1. In both FOG
and local clouds the most massive clumps meet this condition (Fig.~1\,a, b),
while lower-mass clumps have $\alpha >$1 and would need some external pressure
to be confined. Fig.~1\,c, d shows that when the pressure-term due to the
interclump medium is taken into account, virtually all clumps can be brought
in or near equilibrium. 

\begin{figure}            
\plottwo{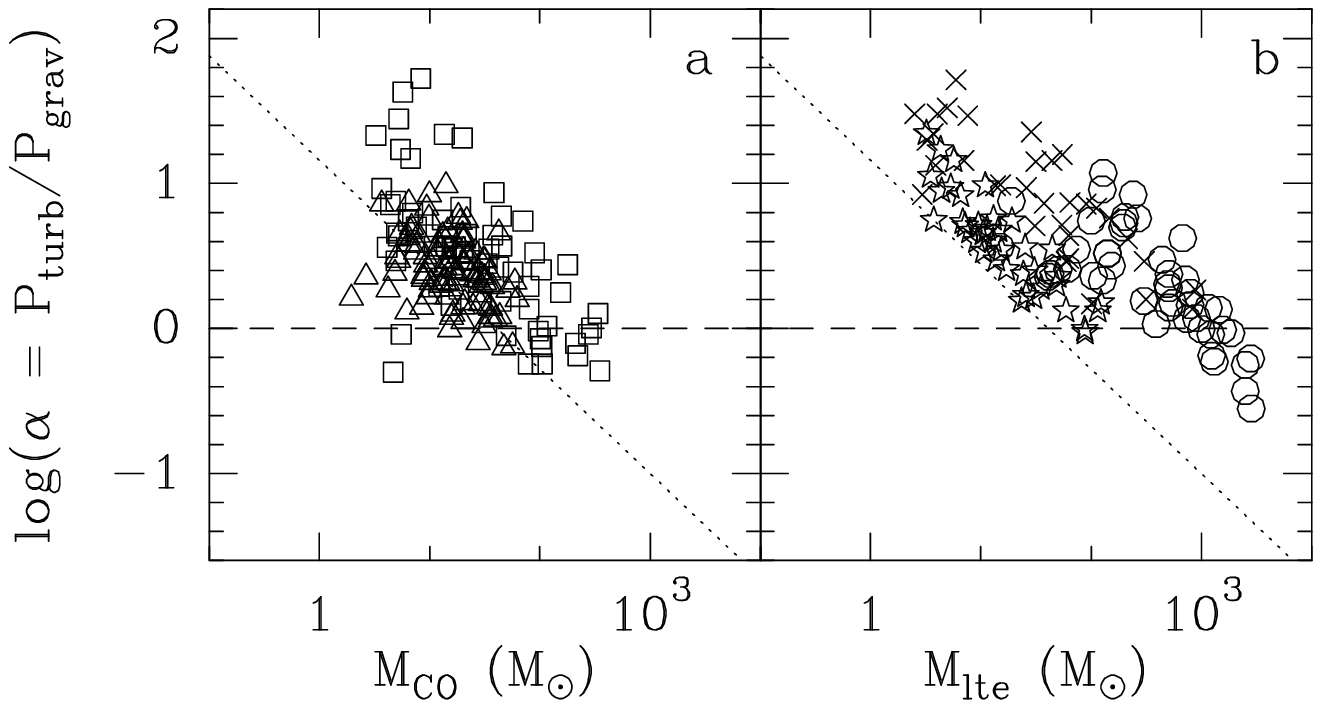}{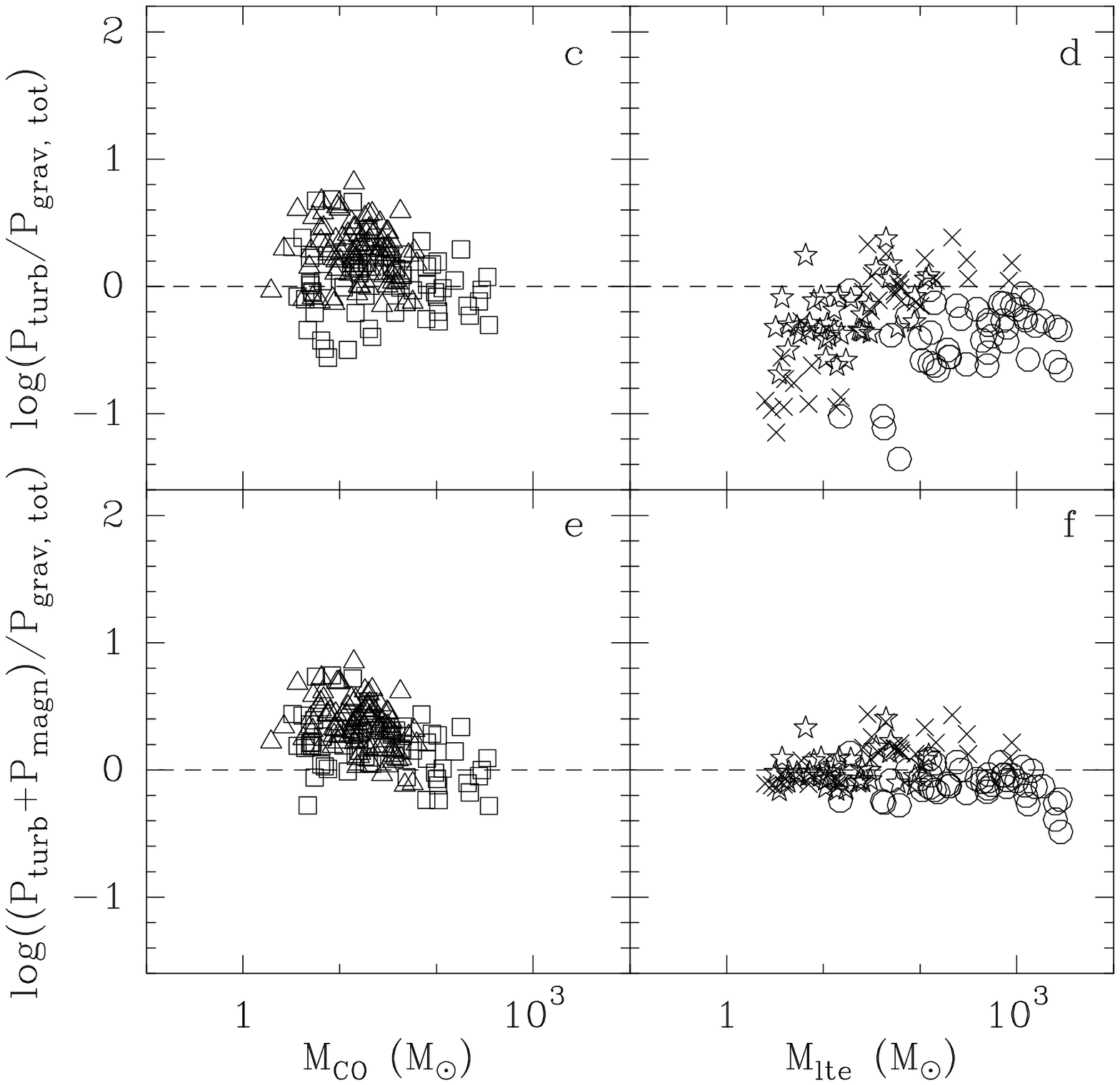}
\caption{
Ratio of the pressures balancing a clump's equilibrium.\hfill\break\noindent
{\bf (a, b)}\ Virial parameter (= ratio of turbulent and gravitational
pressures) as a function of mass for two outer Galaxy clouds (a) and two local
clouds (b).
Symbols are for WB89-85 (open squares), WB89-380 (open triangles), Orion B
South (crosses), and the RMC low-resolution (open circles) and high-resolution
data (stars). The dotted lines are a visual aid to mark the lower envelope of
points in the local sample.\hfill\break\noindent
{\bf (c, d)}\ as (a, b), but now the gravitational pressure is the sum of 
that due
to the clump's self-gravity, and the pressure on the clumps exerted by the
cloud's interclump gas; {\bf (e, f)}\ as (c, d), but the pressure term of a
10~$\mu$G magnetic field has been added to the turbulent pressure.
}
\end{figure}

In fact, for many clumps in the local clouds
(Fig.~1\,d) P$_{\rm turb}$ is more than compensated for by the total
P$_{\rm grav}$, and some stabilizing force is needed to prevent large-scale
collapse. As shown in Fig.~1\,e, f this can be achieved by adding 
the magnetic field-term to the pressure equation. 
The field may be stronger, especially in the more massive clumps and in the
local clouds, while it may be lower in the FOG clouds, but already with the
presently used relatively low value it is clear that the magnetic field is an
important ingredient in the pressure balance.

\section{The importance of self-gravity}
As seen in Fig.~1\,a, b lower-mass
clumps have $\alpha >$1 and would need some external pressure to be confined
(Fig.~1\,c--f). However, in the FOG clouds there are clumps with small values
for $\alpha$ even down to the lowest masses: in the high-resolution RMC data
the lowest-mass clump with $\alpha \le$2 has M$\approx$23.4~M$_{\odot}$, while
this is 2.0~M$_{\odot}$ and 4.7~M$_{\odot}$ in WB89-380 and -85, respectively,
i.e. up to an order of magnitude smaller. Moreover, in WB89-380 there are 14
clumps with M$< 23.4$~M$_{\odot}$ and $\alpha \le$2; in WB89-85 there are 5
clumps that fall within these constraints. In these clumps, gravity is the
dominant force balancing internal turbulence; this implies that gravitational
collapse and star formation may occur more readily even in the smallest of
these outer Galaxy clumps. Since a clump of mass $M$ can only form stars of
mass less than $M$, this would mean that an excess of low mass stars is 
expected to form in the outer Galaxy with respect to local clouds. This 
would provide an explanation for the results of those (Garmany, Conti, \& 
Chiosi 1982; Wouterloot et al. 1995) who find that the IMF steepens in the
outer Galaxy.

\begin{figure}
\plottwo{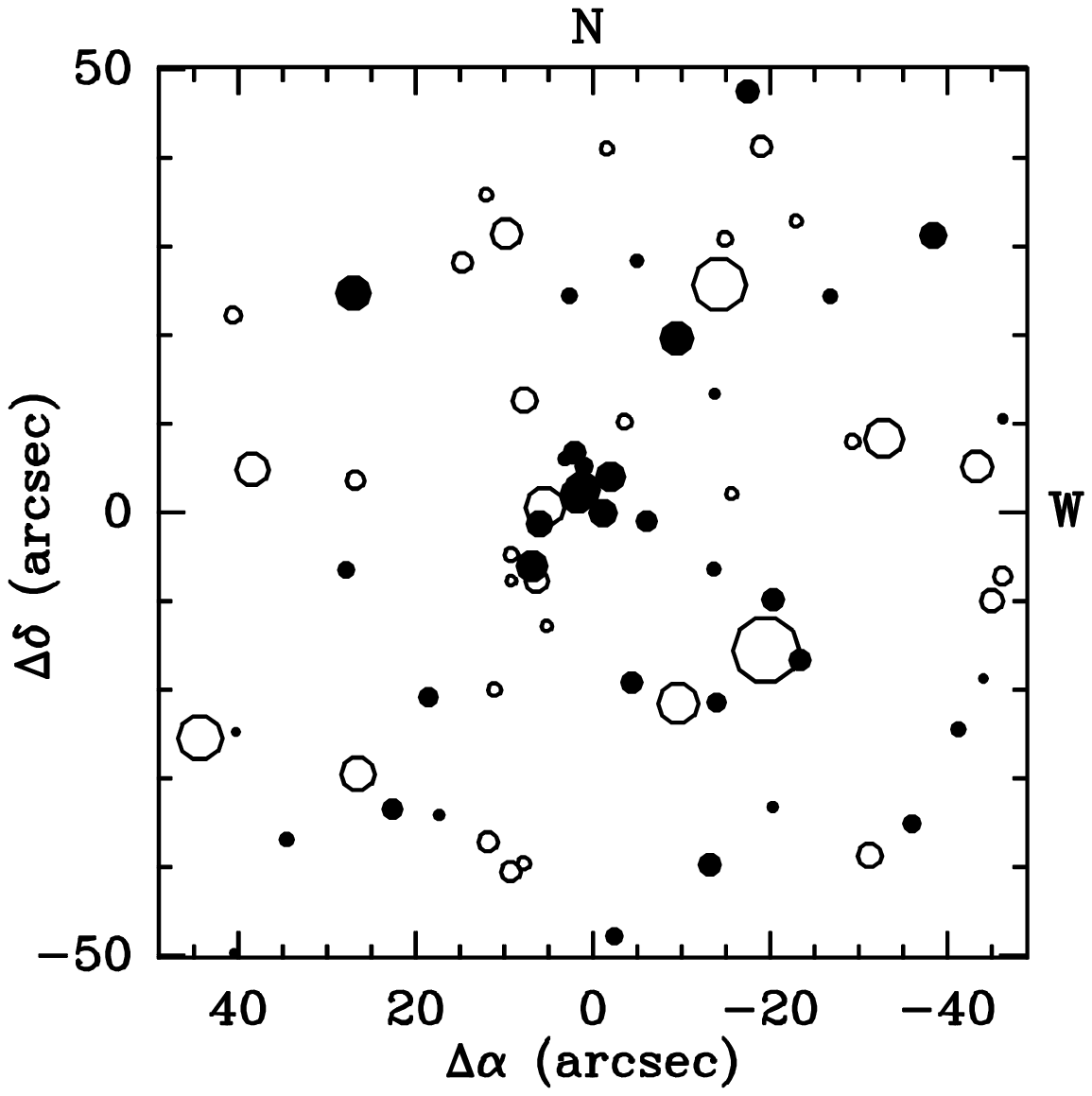}{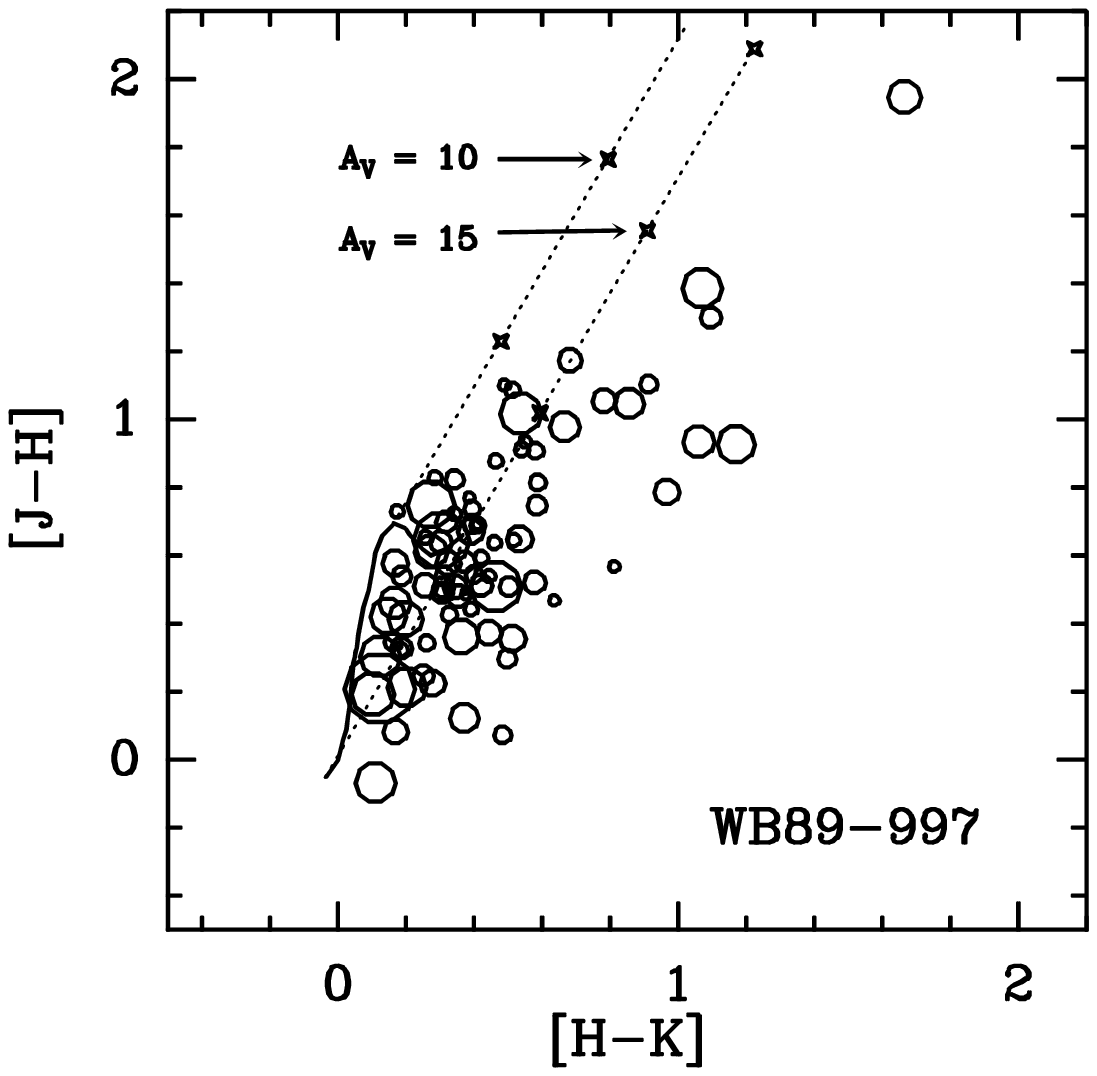}
\caption{Plot (left), based on the NIR (JHK) image of the cluster around
IRAS07257$-$2033 (WB89-997; R, d=15.7, 9.3~kpc; 
Brand \& Wouterloot 2002). 
Size is $\approx 4.5\times 4.5~{\rm pc}^2$. Symbol size is 
proportional to K-magnitude; filled circles indicate stars with intrinsic
IR-excess. On the right is the derived colour-colour diagram.
The size of the circles is proportional to K-magnitude. The drawn curve is the
unreddened main sequence; the dotted lines
indicate the direction of normal interstellar reddening, 
and define the reddening band for normal stellar photospheres. Crosses 
indicate increments of 5~mags of visual extinction.
Objects outside and to the right of this band have intrinsic IR-excess}
\end{figure}

\section{Embedded star clusters}
Do we actually find stars of M$\la$1~M$_{\odot}$ in the far-outer Galaxy, and
are there more of them than what is seen locally? There is indeed ongoing
star-formation in these, and other FOG clouds. Brand \& Wouterloot (2002)
have imaged 5 embedded clusters in 4 clouds in the J, H, K-bands with the
ESO 2.2-m telescope. An example is shown in Fig.~2.
The advantage of observing clusters in distant outer
Galaxy clouds is that that there is negligible contamination by background
stars.
We find stars with visual extinction A$_{\rm V}$ up to about 19 mags. 
The completeness limit of our data is around K$\approx
16.5-17.5$. At a typical distance of 10~kpc and with A$_{\rm V} \sim 10$~mags.
this corresponds to spectral types A0V$-$A5V, i.e. M$\sim 2.9-2$~M$_{\odot}$;
with A$_{\rm V} \sim 20$~mags. this is about B8V, or M$\sim 3.8$~M$_{\odot}$.
To find stars of masses low enough to be formed from the clumps discussed
above, i.e. $0.5-0.8$~~M$_{\odot}$ or types M0V$-$K0V, one needs to go down
to at least K$\approx 21-20$~mags. (at 10~kpc and with A$_{\rm V}$=10), which
is a job for VLT/ISAAC.

\section{Future work}
The results discussed above are necessarily  derived from a rather 
inhomogeneous data base, where effects of resolution, sensitivity and 
isotopomers and transitions play a role. For a detailed discussion of the 
observational biases we refer to Brand et al. (2001). Homogeneous observations 
of a larger sample of outer- and inner Galaxy clouds and their embedded 
stellar population are needed to verify if the difference in mass spectra 
reported here are a truly general phenomenon.

\acknowledgements
We thank Jonathan Williams, Carsten Kramer, and Nicola Schneider for making
their original data available for analysis.

\end{document}